\documentstyle[aps,preprint,eqsecnum]{revtex}
\begin{document}
\draft
\preprint
{$\hbox to 5 truecm{\hfill CGPG-96/1-6}\atop
{\hbox to 5 truecm{\hfill gr-qc/9601037}\atop
}$}
\title{Fundamental thermodynamical equation \\
of a self-gravitating system}
\author{Erik A. Martinez  
\footnote{electronic address:
martinez@phys.psu.edu}} 
\address{Center for Gravitational Physics and Geometry,\\ 
Department of Physics,\\ 
The Pennsylvania State University,\\
University Park, PA 16802-6300, USA}
\maketitle
\begin{abstract}
\noindent{\baselineskip=11pt
The features of the fundamental thermodynamical relation (expressing
entropy as function of state variables)  that arise  from the
self-gravitating character of a system are analyzed.  The models studied
include not only a spherically symmetric hot matter shell with constant
particle number but also a black hole characterized by a general thermal
equation of state.  These examples illustrate the formal structure of
thermodynamics developed by Callen  as applied to a gravitational
configuration as well as the phenomenological manner in which Einstein
equations largely determine the thermodynamical equations of state.
We consider in detail the thermodynamics and quasi-static collapse of a
self-gravitating shell. This includes a discussion of  intrinsic
stability  for a one-parameter family of thermal equations of state and
the interpretation of the Bekenstein bound. The entropy growth associated
with a collapsing sequence of equilibrium states of a shell is computed
under different boundary conditions in the quasi-static approximation and
compared with black hole entropy. Although explicit expressions involve
empirical coefficients, these are constrained by physical
conditions of thermodynamical origin. The absence of a Gibbs-Duhem 
relation and the associated scaling laws for self-gravitating
matter systems are presented.   
\par}   
\end{abstract} 
\vspace{7mm} 
\vfill \eject

\section{Introduction}

Although we expect the  line separating ``gravitational" from
``material" degrees of freedom of a self-gravitating system to
disappear completely in an unified treatment of quantum interactions, it
remains a subject of study at the phenomenological
level. Its exploration may clarify the features of thermodynamics
intrinsic to the gravitational field and in turn  provide us with
useful physical guidance in the search of a quantum description of
spacetime.  In this respect, we propose to revisit an old 
problem  in this paper and evaluate the fundamental thermodynamical
relation for a single component, simple system consisting of a 
spherically symmetric self-gravitating shell at finite temperature.

This investigation allows us to address two related issues.  The first
one concerns the properties of the fundamental equation of a 
system which arise {\it solely} from its
self-gravitating character. In other words, we  desire to restrict as
much as possible the physical form of the entropy function  by using
thermodynamical  arguments based only on the phenomenological
characteristics of the  gravitational field (with only minimal assumptions
about the structure of the matter fields making the system).
As we show below, the content of Einstein equations enters 
into the thermodynamical formalism as the phenomenological gravitational
piece of the thermodynamical equations of state. 
The approach followed in this paper to the fundamental equation of a
system seems therefore to agree in spirit (although from a different
perspective)  with the thermodynamical view of Einstein equations
suggested in Ref. \cite{Ja}.

The second issue concerns the amount of entropy that might arise from the
quasi-static collapse of a matter system. Under certain conditions, a
quasi-static sequence of equilibrium states of a shell can be used to
simulate a realistic physical collapse resulting into a black hole. 
Despite the fact that a
spherical shell possesses no ``gravitational entropy"
\cite{GiHa,Pe,DaFoPa,MaYo}, it is natural to ask how
much of the entropy of the final black hole can be traced to the
(quasi-static) matter entropy of the shell since
not every quasi-static process is reversible.
It is therefore reasonable to calculate the maximum amount of entropy
that might be produced from quasi-static collapse processes obeying
different types of boundary conditions and compare it with black hole 
entropy. The ``remaining" entropy must necessarily have its origin in the
irreversible non-equilibrium late stages of collapse, where a
quasi-static approach breaks down. This type of analysis, besides
having a clear thermodynamical interpretation, does not require a precise
description of complex dynamical processes.

Although some of these questions have been suggested and partially
addressed before \cite{DaFoPa,MaYo,Hi}, we believe that they have not been
satisfactorily answered.  
Davies, Ford, and Page \cite{DaFoPa} considered a 
spherically symmetric black hole surrounded concentrically by a cold
thin shell of matter and found no inconsistency in taking the
gravitational entropy as only that of the black hole. However, they did
not take explicit account of the thermodynamics of the matter
distribution: the shell had only the passive effect of depressing the
temperature of the black hole. Since its state variables were held
fixed, the matter entropy of the shell remained a (negligible) constant.
The limitations of this approach are further discussed in Section IV.
Hiscock \cite{Hi} considered a dust shell collapsing into a
pre-existing black hole and suggested defining its entropy as
(one-fourth of) the difference between the surface areas of the apparent
and event horizons. Although in the context of a dynamical collapse,
this approach did not involve the thermodynamics of the shell itself
and remains an indirect proposal. In Ref. \cite{MaYo} it was shown by
using a path integral representation of a canonical partition function
that the additivity of actions of gravity and matter implies that the
total entropy of a black hole and shell system is the simple sum of the
black hole entropy and the ordinary matter entropy.
Even if the latter was formally identified, 
its explicit dependence on the state
variables could not be determined in that approach.

We wish to employ the present simple model to stress the formal
structure of thermodynamics developed by Callen  \cite{Ca} as applied to a
gravitational system. 
Our strategy in this paper is the following:
We obtain the fundamental equation by direct integration of the first law
of thermodynamics. 
This requires finding the equations of state for the system.
It turns out that a phenomenological consequence of the 
self-gravitational character of the
system is to fix completely its pressure and 
partially its thermal (and chemical) equations of state. 
The fundamental equation can
then be evaluated essentially up to an arbitrary function of the horizon
size $r_+$. 
To specify further the entropy one naturally requires 
an empirical or microscopic  description of the matter fields 
forming the shell. This is not a failure of the model or of our analysis
but a natural  consequence of a thermodynamical treatment.  
Following a procedure common in thermodynamics, we will
adopt the simplest physical choice 
for the undetermined function in the thermal equation of state. 
This choice involves empirical
coefficients characteristic of the matter fields.
Since we desire to illustrate how far one can proceed in an entirely 
phenomenological approach we appeal  to physically reasonable 
macroscopic conditions.
These provide constraints on the fundamental equation of our
system by restricting the values of its empirical coefficients.
Besides assuming the weakest possible restrictions on the matter fields 
(namely, that stress-energies obey the ordinary phenomenological
energy conditions), we select physical equilibrium  states 
as the ones  satisfying the following  properties:  
(1) A particular normalization
for entropy, (2) intrinsic local stability, (3) Bekenstein bound on
entropy and/or validity of the generalized second law of thermodynamics
under quasi-static collapse to a black hole, and (4) the third law of
thermodynamics for matter fields. 
Although these conditions
do not have either a fundamental
character or form a complete set which maximally restricts the
fundamental equation, they are {\it physical} conditions 
of a purely thermodynamical origin which provide insight 
into the structure of physically acceptable fundamental equations.

The thermodynamical results of this paper apply to any gravitational
system whose pressure equation of state as a function of its state
variables possesses the particular particular simple form stated below. 
These systems include
not only a  self-gravitating shell but also a black hole spatially 
bounded by a spherically symmetric surface characterized by a thermal
equation of state which does not necessarily coincide with Hawking's
one.
These slightly generalized  equations of state might be relevant in
studies of quantum corrections to Hawking's formula beyond the
semiclassical approximation.

The paper is organized as follows. We  review briefly in Section II  the 
entropy representation in ordinary thermodynamics and compute 
the gravitational contribution to the entropic fundamental equation. 
The resulting general expression is 
applied to the particular case of a power law thermal equation of state. 
Physical requirements imposed on the fundamental equation are presented
and discussed in detail in Section III.   Quasi-static processes
involving  equilibrium states of the shell are studied in Section IV for
different types of boundary conditions.
For a closed system, the quasi-static motion of the shell is reversible.
Examples of irreversible processes are presented. These results
generalize the results of Ref. \cite{DaFoPa} to configurations that
include explicitly the entropy content of matter. 
Although the  amount of entropy
in a quasi-static shell collapse depends on the precise values of the
empirical coefficients in the thermal equation of state, we
calculate the maximum  values of  the entropy  for a one-parameter
family of  equations of state  and compare them with
the entropy of the resulting black hole.
Finally, we illustrate in Section V the scaling laws for self-gravitating
matter systems and the associated absence of a Gibbs-Duhem relation.
These laws are in clear contrast to the ones
familiar in ordinary flat-space thermodynamics. 
Concluding remarks are presented
in Section VI.  Henceforth we adopt units for which  $c =
k_{\scriptscriptstyle Boltzmann} = 1$, but explicitly display $G$ and
$\hbar$.

\section{Fundamental relation}
 
The fundamental relation of a thermodynamical system in the 
so-called ``entropy representation" expresses the entropy as the function
\cite{Ca}
\begin{equation}
	S = S(M, A, N) \ . \label{fundamental}
\end{equation}
The entropic state variables  of the system are
its proper local energy (denoted here by $M$), its size
(denoted by $A$), and an arbitrary 
number of conserved quantities (denoted generically  by $N$). 
Once known, the fundamental 
equation (\ref{fundamental}) contains {\it all} the  thermodynamical 
information about the system \cite{Ca}. 
In this paper we focus our attention on the entropy representation of
equilibrium states  (as opposed to the alternative energy representation) 
because we are interested in finding the entropy as a function of 
energy and size.

There exist several methods to find (\ref{fundamental}) in ordinary
thermodynamics where self-gravitational effects are considered
negligible. One possible way is by direct integration of  the
first law of thermodynamics 
\begin{equation}
	T \, dS = dM + p \, dA - \mu \, dN
\end{equation}
if one knows the three equations of state 
\begin{eqnarray}
\beta &=& \beta(M, A, N) \ , \nonumber \\
p&=& p(M, A, N) \ ,  \nonumber \\
\mu &=& \mu(M, A, N) \ ,  \label{eos}
\end{eqnarray}
where $\beta =1/T$ denotes the temperature function, $p$ the pressure 
conjugate to $A$, and $\mu$  the chemical potential conjugate to $N$. 
In fact, two equations of state are sufficient to
determine the fundamental  relation up to an undetermined
integration constant \cite{Ca}. Clearly, any single equation of state
contains less information than the fundamental equation.  A second
alternative method in standard thermodynamics  consists in substituting
the three equations of state  (\ref{eos}) in the so-called Euler
relation \cite{Ca}. However, as we will show in Section V, the ordinary
form of the  Euler relation  is not the correct one for a
self-gravitating  matter system. We will use therefore in this paper the
first approach to the fundamental equation and point
thereafter the appropriate form of the Euler relation for the system. 
This approach is technically  simple and, most importantly,
conceptually transparent. Alternative methods which involve calculations
of partition functions or  density of states in terms of functional
integrals \cite{MaYo,BrMaYo,BrYoRev} will not be explored here.

What are the thermodynamical state variables and equations of state 
(\ref{eos}) for a self-gravitating shell in  both thermal and 
mechanical equilibrium with itself?
To answer this question, consider Israel's  massive  thin-shell 
formalism \cite{Is,La}.
As it is well known, an exterior Schwarzschild solution  and an interior
flat solution are joined together across an infinitely thin,
spherically symmetric matter shell. The position of the shell is denoted
by  its circumferential radius $r(\tau$) which is a function of the shell
proper time $\tau$. We  consider in this paper only  equilibrium  
configurations, namely, static (or effectively static)  
configurations in which the shell remains at rest for proper time 
periods much longer than the thermalization  period of the material 
on the shell. The position of the equilibrium configuration is denoted
by $r=R$, and the surface area of the shell by $A  \equiv 4\pi  R^2$.

The junction conditions at the shell require the induced metric 
to be continuous and  the discontinuity in the extrinsic curvature  to
be proportional to  the stress-energy tensor in this hypersurface
\cite{Is}. The latter can be decomposed in
terms of a surface energy  density $\sigma$ and a surface pressure  $p$. 
The proper, locally  defined mass $M$ of the shell 
is related to  the surface energy density $\sigma$ by 
$M \equiv 4 \pi R^2 \sigma$. 
The junction conditions  imply that the ADM energy $m$ is 
given  in terms of the energy $M$ and position $R$ as \cite{Is,DaFoPa} 
\begin{equation}
	m(M,R) = M  - {{GM^2}\over{2R}}  \ .\label{m}
\end{equation}
The ADM energy is the sum of the proper energy $M$ and the gravitational
binding energy associated with building the shell \cite{Yo,DaFoPa,MaYo}.
This equation can be rewritten in the familiar form
\begin{equation}
	M = {R \over G} \, (1 - k) \ , \label{M}
\end{equation}
where it is useful to introduce the notation
$k \equiv  (1 - r_{+}/R)^{1/2}$. 
The quantity $r_+ (M,R) \equiv 2 G m(M,R)$  denotes the Schwarzschild 
radius of the  shell.
The junction conditions also determine the value of the equilibrium 
pressure.
For a shell to be effectively static its tangential pressure must have 
the precise form  
\begin{equation}
	p(M, R) = {{GM^2}\over{16 \pi R^3}} 
\Bigg(1 - {{GM}\over{R}}\Bigg)^{-1}  = 
\, {{1}\over{16 \pi \, G \, R \, k}} (1 -k)^{2} \ .\label{p} 
\end{equation}

The thermodynamical state variables for the system in the entropy
representation are  the local energy $M$,  the surface area 
$A \equiv 4 \pi R^2$, and the conserved number $N$. 
(Because of spherical symmetry, we use $R$ and $A$ 
interchangeably in what follows.)
We will assume throughout the equilibrium surface energy 
density $\sigma$ and  pressure $p$ to be non-negative. 
The state variables $(M, R)$ are therefore 
non-negative. We will also assume that 
$R \geq r_{+} \geq {l_p}$, where ${l_p} = 
(G\hbar)^{1/2}$ denotes the  Planck length. In particular, 
this implies that the
thermodynamic state space is such that  $0 \leq k \leq 1$. 

Since both $\sigma$ and $p$ are  non-negative, the shell matter 
automatically satisfies  the weak energy  and time-like convergence 
conditions \cite{HaEl,FrHoKo}. It is well known that the dominant energy 
condition $p \leq \sigma$ further constraints the position of the shell 
to obey $R \geq 25/24 \, r_+$, or equivalently,  
$k \geq 1/5$.

We wish to evaluate the entropy using a minimal set of assumptions about
the matter fields making the shell. Consequently,  we focus attention in
the case when the number of particles $N$ is constant and ignore the form
of the chemical potential.  We therefore require only two equations of 
state to determine the fundamental relation  up to an 
additive constant \cite{Ca}. Observe that
Eq. (\ref{p}) does indeed provide  the desired pressure equation of state
for  the shell.  We emphasize  that this equation is 
a {\it unique} consequence of the gravitational equations 
across the shell hypersurface and is independent of the nature 
of the matter fields making the shell.

Consider now the first law of thermodynamics for a hot shell 
whenever the total number of particles $N$  is   constant. By
virtue of the pressure equation of state (\ref{p})  and the differential
form of the local energy expression (\ref{M}) the first law can be
suggestively written as a total differential of the form:   
\begin{eqnarray}
	T \, dS 
	&=& dM +  p \, dA \nonumber \\
	&=&{{1}\over{2Gk}} \,\, dr_{+} \ , \label{TdS}
\end{eqnarray}
where $T \equiv 1/\beta$ denotes the local temperature of the shell at
the equilibrium position $r=R$.   
The result (\ref{TdS}) is non-trivial despite its 
familiar form.
The shell  possesses a local mass $M \neq m$, and  a non-zero 
pressure $p$ which keeps it static at an equilibrium position. 
The identification of $-p \, dA$ as mechanical work and $T \, dS$ as heat
transfer refer to quasi-static processes of the shell itself. 
The particular form (\ref{TdS}) of the first law is a consequence of  the
pressure equation of state (\ref{p}), and therefore, of the 
gravitational junction conditions.

The local temperature $T$ appears in the first law as an 
integrating factor. The integrability condition for the entropy $S$
provides an equation for $\beta (M,R)$ of the form 
\begin{equation}
	\Bigg({{\partial \beta}\over{\partial A}} \Bigg)_M = 
	\Bigg({{\partial \beta p}\over{\partial M}}\Bigg)_{A}  \ ,
\label{intcond}
\end{equation}
with the pressure given by Eq. (\ref{p}). 
Under a  change of variables from $(M,R)$ to 
$(r_{+}, R)$  the
integrability equation becomes
\begin{equation}
	\Bigg({{\partial \beta}\over{\partial R}}
	 \Bigg)_{r_{+}} = 
	{{\beta}\over{2\, R \, k^2}} \, \big(1 - k^2 \big) \ ,
\end{equation}
whose general solution  is
\begin{equation}
	\beta (M, R) =  b(r_{+}) \, k \label{beta} \ ,
\end{equation}
where $b(r_{+})$ is an arbitrary function of the
quantity $r_{+}(M, R)$.  The function $b(r_+)$ can
therefore be interpreted as the
inverse temperature the shell would possess {\it if} located at
spatial infinity. Equation (\ref{beta}) is a consequence of the 
integrability conditions for entropy and naturally represents the 
equivalence principle \cite{To} as applied to a self-gravitating system
at non-zero temperature. 
While the integrability condition forces the function $b$ to be 
dependent on the state variables  through the quantity $r_+(M,R)$, it
does not determine its precise dependence. 
This is physically reasonable and expected from other grounds: in a
path integral description of the partition function for the system, the
flat spacetime geometry in the region inside the shell can be
periodically identified with any proper period \cite{MaYo}.

Substitution of the inverse temperature (\ref{beta}) into the first law
(\ref{TdS}) implies  
\begin{equation}
	dS = {{1}\over{2G}} \, b(r_+) \, dr_{+} \ . \label{dS}
\end{equation}
The fundamental equation for the system is therefore
\begin{equation}
	S(M, R) =  {{1}\over{2G}} \int b(r_{+}) \, dr_{+} 
+ \, S_0 \ , \label{egral}
\end{equation}
where $S_0$ is an integration constant. Notice that the
entropy is a function of the state variables $(M,R)$ only through the
quantity  $r_+ (M, R)$. In general, the quantity $S_0$ is only a function
of the number of particles $N$. Since the latter is constant in our
model, the quantity $S_0$ is  a number.

The entropy expression (\ref{egral}) is a consequence of the 
self-gravitating character of the model and constitutes one of the main
results of this section.  It follows directly from the gravitational
junction conditions (\ref{m}) and (\ref{p})  and the equivalence principle
(\ref{beta}).  The former determine the pressure equation of state
whereas the latter determines the redshift factor in the temperature
equation of state. (As is expected  the equivalence principle  also
determines the redshift factor  in the chemical potential equation of
state \cite{MaYo}.) 
Equations (\ref{beta}) and (\ref{egral}) apply to  {\it every}
self-gravitating shell with $N = {\rm const.}$ independently of its
matter composition. 
A concrete form of 
the function $b(r_+)$ in the fundamental equation has to originate
in an explicit model of the matter fields.

The calculation of the entropy  (\ref{egral}) is clearly reminiscent of
the  calculation of black hole entropy.
This is so because the quasilocal energy $E$ and pressure $s$
characteristic of a  Schwarzschild geometry of ADM mass ${\tilde m} =
{{\tilde r}_+}/2G$ enclosed inside a  boundary surface of radius $r_0$
are respectively \cite{Yo,MaYo,BrYo1} 
\begin{eqnarray}
E &=&{{r_0} \over G} \, (1 -  {\tilde k})\ , \label{Ebh} \\  
s &=&{{1}\over{16 \pi \, G \, {r_0} \, {\tilde k}}} \, 
(1 -{\tilde k})^{2} \ , \label{pbh}
\end{eqnarray}
where ${\tilde k} \equiv (1 - {{\tilde r}_+}/{r_0})^{1/2}$.
Both quasilocal energy (\ref{Ebh}) and pressure (\ref{pbh}) 
associated to  a
black hole are defined in terms of the two-dimensional surface that 
contains the system, and possess the same functional form as expressions 
(\ref{M}) and (\ref{p}) for a shell. 
However, the thermal equation of state
for a  black hole in thermal equilibrium with a heat bath is given
uniquely in the semiclassical approximation by Hawking's temperature
formula \cite{Ha}
\begin{equation}
\beta_{H} (r_0) =  {{4 \pi G {\tilde r}_+} \over {{{l_p}}^2}} 
\,\, {\tilde k} \ . 
\label{betabh}
\end{equation}
The integration left undone in (\ref{egral}) can  be carried out
explicitly for a black hole, yielding the well-known Bekenstein-Hawking
formula
\begin{equation}
S_{BH} = \pi \Bigg( {{{\tilde r}_+}\over{{l_p}}} \Bigg)^2\ , \label{Sbh}
\end{equation}
where the entropy is normalized to zero for a zero mass black hole.

\subsection{Power law equation of state}

Consider now the simplest possible choice for the function $b(r_+)$ in
the thermal equation of state (\ref{beta}). This is clearly a  power law
expression of the form  
\begin{equation}
b(r_{+}; \eta, a) = {{2 \, G \,\eta}\over {{l_p}^{(1+a)}}} \,\, 
{r_+}^{a} \,  
 \ , \label{powerlaw}
\end{equation}
where $\eta$ and $a$ are two empirical coefficients that characterize
the matter fields in the shell. We treat $a$ as a real
parameter. For simplicity and with no loss of generality, we consider
$\eta$ as a dimentionless number and display the units explicitly in 
the previous equation. 
In later sections we will pay special attention to  the case when 
$\eta$ is of order one.
This condition  simplifies the model and allows to study the contribution
of the terms involving $r_+$ in order of  magnitude estimates of the shell
entropy.

By substituting (\ref{powerlaw}) in (\ref{beta}) we obtain the simplest 
one-parameter family of thermal equations of state. Positivity and
finiteness of temperature imply that $\eta $ is non-negative.  
With this choice for the thermal equation of state the entropy  becomes
\begin{equation}
	S(M, R; \eta, a) = {{\eta}\over{(a+1)}} 
	\Bigg( {{r_{+}}\over{{l_p}}} \Bigg)^{(a+1)} 
	+ \, S_0  \label{Sresult}
\end{equation}
for parameter values $a \neq -1$, and 
\begin{equation}
S(M, R; \eta) = {\eta} \ln (r_{+}) + S_0 \label{Sa-1}
\end{equation}
in the case $a = -1$. In our model $S_0$ is itself a numerical
constant. Observe that  for $a=1$ the thermal equation of state is linear
in $r_+$. 
In this case the entropy has the same functional dependence 
on $r_+$ as the Bekenstein-Hawking black hole entropy
(\ref{Sbh}). In contrast, the temperature equation of state is
independent of the extensive variables $(M,R)$ in the case $a=0$.

\section{Physical constraints} 

Thermodynamics alone cannot fix uniquely either the empirical
coefficients $\eta$ and $S_0$ or the empirical parameter $a$.
Their precise values must necessarily arise from a 
description of the micro-physics of the physical shell. 
However, the next step in our  approach is to investigate
conditions which {\it physical} equilibrium states of the system must
necessarily satisfy.  We focus  attention on the  restrictions  imposed 
phenomenologically by these conditions on the range of values of 
the empirical coefficients appearing in the fundamental equation 
(\ref{Sresult}).

\subsection{Normalization of entropy}

The entropy can  be defined up to an absolute constant.
However, it seems physically reasonable to assume that a zero mass
shell must possess zero entropy.   By Eq. (\ref{egral}), this
condition  restricts the  area under the function $b(r_+)$
in the limit of zero mass, namely
\begin{equation}
\int b(r_+)\, dr_+ + S_0  \to 0 \quad as \quad M \to 0 \ . \label{norm}
\end{equation}
Consider the power law thermal equation of state (\ref{powerlaw}) and 
the
entropy  (\ref{Sresult}). 
Since the quantity $r_+$  also vanishes 
if the proper mass $M$ vanishes, the above condition  is
satisfied for  $a > -1$ whenever  $S_0 = 0$. 
In contrast, the entropy diverges for $a \leq -1$
as  $M$  tends to zero for any finite value of  $S_0$. This is
clearly not physical.  The normalization condition therefore constrains 
the coefficients to be  $a > -1$ and $S_0 =0$.

\subsection{Intrinsic Stability}

We can study the intrinsic stability of the thermodynamic equilibrium
states by  direct inspection of the fundamental relation. 
Global stability in the entropy representation requires that the entropy 
hypersurface  lies everywhere below its tangent two-dimensional planes 
\cite{Ca}.  We focus our attention here on the local intrinsic stability
conditions which, although weaker than the concavity of the entropy
stated above, insure that the entropy function does not  increase due to
inhomogeneities of the state variables. 
(Stronger stability criteria for a shell which may or
may not overlap with  the one adopted here are briefly discussed in the
concluding section.)
Since $N$ is assumed constant, we deal only with a three-dimensional 
thermodynamic space defined by the variables $(S, M, R)$. 
In terms of the fundamental
equation, local  intrinsic stability is guaranteed if the following 
three  inequalities are satisfied simultaneously, namely
\begin{eqnarray}
&\Bigg( & {{{\partial}^2 S}\over{\partial M^2}} \Bigg)_A \leq  0 \ , 
\label{sta1} \\
&\Bigg( & {{{\partial}^2 S}\over{\partial A^2}} \Bigg)_M \leq  0 \ , 
\label{sta2} \\
&\Bigg( & {{{\partial}^2 S}\over{\partial M^2}} \Bigg) 	
\Bigg({{{\partial}^2 S}\over{\partial A^2}} \Bigg) - 
\Bigg({{{\partial}^2 S}\over{\partial M \partial A}} \Bigg)^2 {\geq} \, 0
\ .  \label{sta3}
\end{eqnarray}
Conditions (\ref{sta1}) and (\ref{sta2}) insure that the intersection of
the entropy surface with  planes of constant $M$ or  $A$ have
negative curvature, whereas the ``fluting" condition (\ref{sta3})
insures the equivalent under coupled inhomogeneities of $M$ and $A$.

Although the criteria (\ref{sta1})-(\ref{sta3}) can be expressed as a set
of differential  inequalities for the function  $b$, 
we do not write these explicitly. 
Instead, consider them  as applied to the fundamental equation 
(\ref{Sresult})-(\ref{Sa-1}). 
Our goal is to find the stability regions 
in the thermodynamical state space $(M,R)$ of the shell as a function of 
the parameter $a$. Consider first the case $a \neq -1$.
It is easy to see that the condition (\ref{sta1}) is
automatically satisfied by any physical $k$ (that is, in the range 
$0 \leq k \leq 1$) if $a \leq 0$, whereas it is
satisfied for  $a>0$ provided 
\begin{equation}
	k \leq \sqrt{{{1}\over{2a + 1}}} \ , \label{ineq1}
\end{equation}
or equivalently, if $R \leq (1 + 1/2a) r_{+}$.  
Notice that the dominant  energy condition
requires  $k \geq 1/5 $.  Therefore, the stability condition
(\ref{sta1})  and the dominant energy condition jointly restrict 
the value of the parameter $a$ to be $a \leq 12$.

Condition (\ref{sta2}) is satisfied automatically by any  
$0 \leq k \leq 1$ if $a \leq 3$, whereas for 
$a \geq 3$ it requires 
\begin{equation}
	k \geq 1 - {6\over{(a+3)}} \ , \label{ineq2}
\end{equation}
or equivalently, that $R \geq (a+3)^2 /(12a) \, r_{+}$.  
Observe that for parameter values $a > 3 + \sqrt{6} \approx 5.45$, 
the two conditions  (\ref{sta1}) and (\ref{sta2}) cannot be satisfied
simultaneously. 

Consider now the ``fluting" condition  (\ref{sta3}).
It is not difficult to show that it implies the inequality
\begin{equation}
	 (3a + 1) k^2 +  2 a k + (a - 1) \leq 0 \ . \label{ineq3}
\end{equation}
This relation can not be satisfied for parameter values $a\geq 1$ if
$k$ is in the  range $0 \leq k \leq 1$. (For $a=1$ the inequality is 
marginally satisfied at  $k=0$, or equivalently, at $R= r_+$.)
The left-hand side of (\ref{ineq3}) is  always smaller
than  $6a$. This implies that the inequality  is 
automatically satisfied for $a \leq 0$. 
The non-trivial range to analyze is therefore $1 > a > 0$.
For values of $a$ slightly larger than zero, Eq. (\ref{ineq3})
restricts the 
values of $k$ to be slightly smaller than one, whereas for values of $a$ 
slightly smaller than one, the inequality restricts $k$ to be slightly 
larger than zero.  For example, if $a= 1/3$, $k \leq 0.44$ 
($R \leq  1.24 \, r_+ $), whereas if $a=2/3$, $k \leq 0.175 $ 
($R \leq  1.03 \, r_+ $). 
Interestingly, the inequality (\ref{ineq3}) implies $k = 1/5 $ for the
particular parameter value 
\begin{equation}
	a = {{12}\over{19}} \ .
\end{equation}
For values of $a > (12/19) $, $k$ is restricted
to take values smaller than $1/5$. 
This clearly contradicts the dominant energy condition. 
Therefore a shell may satisfy simultaneously the stability 
condition (\ref{sta3})  and the dominant energy condition 
provided $a \leq 12/19$. 
This restriction on the  parameter $a$ is more stringent than the ones 
implied by  conditions (\ref{sta1}) and (\ref{sta2}).
In fact,  condition (\ref{ineq1}) is superseded by  
(\ref{ineq3}): it can be shown that  (\ref{ineq1}) is automatically
satisfied if (\ref{ineq3}) is satisfied, whereas the inverse is not true.

For completeness, consider the case $a=-1$. 
The stability criteria are  automatically satisfied by the entropy
formula (\ref{Sa-1}) for all values of the  state variables satisfying 
$R \geq r_+ $. In particular, the condition (\ref{sta3}) implies 
$k^2 + k + 1 \geq 0$,  which is valid for any physical $k$. 

As mentioned before, one recovers the Bekenstein-Hawking expression
for entropy in the limit $a=1$ and $\eta = 2\pi$ (``black hole case"). 
The previous analysis shows that a system characterized by this
fundamental equation is {\it not} intrinsically stable. 
This has been noted in the context of statistical ensembles in Refs.
\cite{ensembles}. 

Table 1 illustrates the stability regions for different values of the 
parameter $a$. Summarizing, the stability criteria 
(\ref{sta1})-(\ref{sta3}) 
do restrict the range of the exponent $a$ to be $ -\infty < a < 1$. 
In the parameter  range  $-\infty < a \leq 0$ the shell is 
intrinsically stable for 
{\it every} position $R \geq 25/24 \, r_+$. 
In the range $0 <a < 1$ stability restricts the possible values of 
its position  according to Eq. (\ref{ineq3}).
Intrinsic local stability together with the normalization (\ref{norm}) of
the entropy  
necessarily imply  $ -1 < a < 1$.
Finally, the dominant energy condition further restricts the 
range to
\begin{equation}
-1 < a \leq \, {{12}\over{19}} \ . \label{finalrange}
\end{equation} 
Only for these values of the parameter $a$ can an 
intrinsically  {\it stable} shell be located at a position 
 $(25/24)\, r_{+} \leq R < \infty$.  
The smaller the value of $a$, the
larger  the range of possible radii $R$ for the shell.

We close this subsection with a final remark. The result 
(\ref{finalrange}) implies that the exponent of $r_+$ in the entropy
formula  (\ref{Sresult}) for an intrinsically stable shell obeying 
energy  conditions has an upper limit 
$(a+1) \leq (31/19) \approx 1.63 < 2$.

\subsection{Second Law and Bekenstein bound}

A quasi-static sequence of equilibrium states 
describing a collapsing shell is described in detail in the following 
section. Assuming that a black hole forms as the end-point of
this process, the generalized second law of thermodynamics  would require
the entropy $S_{BH}$ of  the black hole to be larger or equal than  
the entropy $S$ of the shell as it crosses its own horizon. 
We will not
discuss in this paper whether the  second law is satisfied 
because of  buoyancy forces of the type discussed in  Refs. \cite{UnWa} 
or because of a fundamental bound in entropy of the type introduced 
in  Refs. \cite{Be81,Be94}. 
In the spirit of thermodynamics, we  wish only to emphasize the
restrictions it imposes on thermodynamical parameters and
discuss the interpretation of the thermodynamic quantities involved in
the Bekenstein bound for the system.
For a shell obeying the fundamental equation (\ref{Sresult}) the 
 second law will  be  satisfied provided 
\begin{equation} 
S = {{\eta}\over{(a+1)}} 
\Bigg( {{r_{+}}\over{{l_p}}} \Bigg)^{(a+1)}  \leq 	
\, S_{BH} = \pi \Bigg( {{r_{+}}\over{{l_p}}} \Bigg)^2 \ ,
\label{seclaw}
\end{equation}
where we have assumed $S_0 =0$ and $a > -1$.  
This in turn restricts the numerical value of the coefficient $\eta$
to be 
\begin{equation}
\eta \leq \pi \, (a+1)  \Bigg( {{r_{+}}\over{{l_p}}}
\Bigg)^{(1-a)} \ . \label{eta1}
\end{equation} 
The second law can be satisfied with  values for  
$\eta$ not necessarily of order one.  However, in our general model
(\ref{powerlaw}) the coefficient $\eta$ was assumed to be a number
and not a function dependent on the quantity $r_+$. For simplicity,
its value must remain unchanged for every choice
of $r_+$, either in the case of $r_+$ being constant (closed system) or 
in the case when $r_+$ varies in an open quasi-static collapse. 
(These processes are discussed  in Section IV.) In particular, we
desire to satisfy (\ref{eta1}) for all  $r_{+} \geq \, {l_p}$ with a
single value of  $\eta$.  A sufficient (but not
necessary) condition that guarantees the second law and the above
requirements is obtained by taking the infimum of the right-hand side of
(\ref{eta1}).  Since $(1-a)\geq 0$,  this implies 
\begin{equation} 
	\eta \leq \pi \, (a+1) \ . \label{eta2}  
\end{equation}
This inequality is perhaps too restrictive in general since for 
given  $r_+$ the second law can be satisfied for larger values of
$\eta$. However, as already mentioned, it allows us to illustrate in
the next section the order of magnitude of the contribution to the  
entropy (\ref{Sresult}) arising from powers  of $r_+$.

Notice that Eq.  (\ref{eta1}) restricts the values of
temperature.  Substitution  into (\ref{powerlaw}) yields 
\begin{equation}
T_\infty \geq T_c \equiv {{2}\over{(a+1)}}\, T_H \ ,
\end{equation}
where $T_{\infty}$ is the temperature the shell would have  if
located at spatial infinity, and 
$T_{H} \equiv {\hbar}/(4 \pi r_{+})$
denotes Hawking's asymptotic temperature. It appears  therefore
that for asymptotic values of the temperature of the shell {\it smaller} 
than the critical value $T_c$ the generalized second law would be
violated. Apparent paradoxes of this type have appeared in other systems 
(see for example Ref. \cite{Be73}). 
They indicate that in this regime  statistical fluctuations do become 
dominant and a thermodynamical description is  inappropriate. 
Nevertheless, in our model we assume Eq. (\ref{eta2}) as the
defining value of $\eta$. This implies that the local
temperature is such that
\begin{equation}
T_\infty \geq  \, {{2}\over{(a+1)}} \, \bigg( {{r_+}\over{{l_p}}}
\bigg)^{(1-a)} \, T_H \ .
\end{equation}
Since $a <1$, the temperature remains much larger than $T_c$  for
macroscopic shell configurations for which $r_+ \gg \, {l_p}$.

Consider now the entropy bound proposed by Bekenstein \cite{Be81,Be94}
in the 
particular case of a self-gravitating shell. For an object of maximal
radius  ${\cal R}$ and total energy ${\cal E}$ the proposed bound reads 
\begin{equation}
S \leq {{2\pi {\cal R} {\cal E}}\over\hbar} \ . \label{bb}
\end{equation} 
In our model the bound restricts the area under the function $b(r_+)$:
\begin{equation}
\int b(r_+) dr_+ \leq {{2\pi {\cal R} {\cal E}}\over\hbar} - S_0 \ .
\end{equation} 
How are the  quantities ${\cal R}$ and ${\cal E}$ to be interpreted in
this case? It seems natural to assume ${\cal R} = R =
(A/4\pi)^{1/2}$, since the latter is a unique measure of the size of
the shell. There are however at least two possibilities for
interpreting ${\cal E}$:  either as the local mass $M$ or as the ADM mass
$m$.  If ${\cal E}$ is interpreted as $M$, the bound implies  
\begin{equation}
S \leq {2\pi} \, \bigg({{R^2}\over{{{l_p}}^2}} \bigg) \, (1 - k) \ .
\label{bb1} 
\end{equation}
As the shell crosses its own horizon ($R \to r_+$), the right hand side 
of (\ref{bb1}) tends to $2 S_{BH}$.   
On the other hand, if ${\cal E}$ is interpreted as $m$, the bound implies
\begin{equation}
S \leq {\pi}  \bigg({{R \, r_+}\over{{{l_p}}^2}} \bigg) \ .
\label{bb2}
\end{equation}
As the shell crosses its horizon, the right hand side of (\ref{bb2})
tends to $S_{BH}$. Hence, if a black hole forms  in the limit $R =r_+$ 
of a quasi-static collapse, the Bekenstein bound (\ref{bb}) guarantees 
the validity of the second law if the quantity  ${\cal E}$ is 
interpreted {\it not} as the local proper energy $M$ but as the
ADM energy $m$.  This is simply because, as $R \to r_+$,  
$M \to 2m$ and not to $m$, as can be seen from Eq. (\ref{M}). The bound
that guarantees the second law  can be written in terms of the local
state variables  $(M,R)$ as 
\begin{equation}
S \leq {{2\pi R M}\over\hbar} \bigg(1 - {M \over{2R}} \bigg)  \ .
\end{equation}

\subsection{Third law and Summary}

In its simplest form, the third law of standard thermodynamics  
requires the entropy  to vanish in  the state of zero temperature. 
From Eq. (\ref{beta}), the latter occurs for our system
whenever the function $b(r_+)$ diverges. If this function has the power
law dependence (\ref{powerlaw}) it will diverge for $a > 0$ if  $r_{+}$
diverges, though in this  state the entropy  (\ref{Sresult})
diverges (assuming finite $\eta$). For $a \leq -1$, $b$ diverges as 
$r_{+}$ tends to zero, but in  this state the entropy 
 diverges as well. In contrast, for  $-1 <a <0$ the function $b$ diverges
and the 
entropy vanishes as $r_+$ tends to zero. 
Notice that the temperature cannot be zero for the case $a=0$.
If the above mentioned states could be reached by a
quasi-static  sequence of thermodynamic equilibrium states,
the third law in the above mentioned form would further restrict 
the parameter 
$a$ to the range $-1 <a \leq 0$. 

However, we do not consider this restriction fundamental.
Firstly, it is not a basic postulate in ordinary thermodynamics. It is
not  unusual to encounter reasonable fundamental  equations in 
thermodynamics which do not 
satisfy the third law (an example is the ideal
van der Waals fundamental equation \cite{Ca}).
Secondly,
it is not clear if the formulation of the third law used above 
is the correct one for self-gravitating matter. In particular,
it does not apply to  black holes, where alternative versions exist
\cite{Wa}. 
Finally, the state of zero temperature may not be reachable by
a finite number of  quasi-static equilibrium states of a macroscopic 
shell. Since the thermodynamical treatment breaks down in the
limit $r_+ \approx {l_p}$ due to  quantum gravitational fluctuations
becoming non-negligible,  states for which  $r_+ = 0$ cannot be reached
by a finite number of steps within the present approximation. Therefore,
a violation of the third law implies at most  that the fundamental
equation is not a very good approximation  at very low temperatures.

The results of this  section can be summarized as follows. 
For a power law thermal equation of state,
the fundamental equation for an intrinsically stable shell is 
\begin{equation}
S(M, R; a) =  \pi
\Bigg( {{r_+}\over{{l_p}}}  \Bigg)^{(a+1)} \ , \label{Sfinal}
\end{equation}
where $r_+ (M,R)$ is given by Eq. (\ref{m}) and $-1 < a \leq \, 12/19$.
We have adopted in (\ref{Sfinal}) a value for the coefficient $\eta$
of order one which respects the generalized second law. If
enforced, the third law in its standard form may further restrict $a$ to
be smaller or equal than zero.

\section{Quasi-static collapse and maximum entropy}

A quasi-static process consists of a dense succession of
equilibrium states \cite{Ca}. 
There are no obstacles in principle in constructing  a quasi-static
sequence of equilibrium states of a shell that simulates at least part
of its dynamical collapse.   One can imagine infinitesimal differences
between the pressure of the shell and the gravitational pull which will
force the shell to  collapse gradually.  A quasi-static
sequence can be  expected to be a good approximation to the true dynamical
process only if the time of thermalization of the shell with itself is
relatively small compared to the characteristic times of collapse. 
The thermalization time depends on the material in the shell.
However, it is physically  natural to expect a good agreement for shell 
radii large compared with the horizon
radius. It is not unreasonable to assume that this
approximation  breaks down for  radii in the neighborhood (if not
larger) of the minimal radius $R = 25/24 \, r_+$ at which the dominant
energy condition is marginally satisfied.  In this section we will not be
interested in the precise distances at which the approximation becomes
unphysical but assume the maximal possible range of positions,
namely $25/24 \, r_+ \leq R < \infty $.

Quasi-static processes can be used to simulate a real dynamical process 
in a closed system only if the total entropy is a non-decreasing function 
along the process \cite{Ca}. Thus, the processes can be  either
reversible (if the quasi-static increase of entropy $dS$ 
along the process is zero) or irreversible (if the increase of entropy 
$dS$ along the process is positive). In fact, although every reversible 
process coincides with a quasi-static process, not every quasi-static 
sequence is reversible.
It is natural to try to find the maximum amount of entropy whose origin
can be ascribed to a 
 quasi-static collapse processes and compare 
it with the entropy of the latter. As mentioned in the Introduction, the
``remaining" entropy must  have its  origin in the irreversible
non-equilibrium stages of collapse.

Consider the fundamental equation (\ref{Sresult}). 
It implies that a quasi-static  sequence of shell configurations for
which  $r_+ = {\rm const.}$ is reversible, whereas a quasi-static
sequence for which $r_+$ increases is  irreversible. (Throughout
this paper we refer only to quasi-static processes for a simple, single
shell as opposed to processes for composite systems which could include a
shell in interaction with a heat bath.)
The reversible process is constructed with  
equilibrium  states whose extensive variables $(M,R)$ obey Eq.
(\ref{m}) with fixed constant $m$. 
It is not difficult to prove that
the condition $r_+ = {\rm const.}$ defines a closed system: 
If the shell is imagined located  in a finite region bounded by a
surface whose radius is larger than $R$, the above condition is a
consequence of fixing both the quasilocal energy contained inside the
surface that bounds the system and its size \cite{MaYo}.
Therefore, if the system is closed, the motion of a shell is reversible.

We emphasize that the above results are not trivial and are not included
in the results presented in Ref. \cite{DaFoPa}. In our case the state
variables of the shell are allowed to vary in a quasi-static manner. For
variations respecting the constraint $r_+ = {\rm const.}$, the
quasi-static motion turns out to reversible. In Ref. \cite{DaFoPa} the
state variables of the shell were kept ``frozen" during a quasi-static
variation of the internal black hole parameters. Naturally the matter
entropy remained a negligible constant. This led to the conclusion that
the motion of a shell is always reversible. This was only so because the
matter entropy was not included in the analysis. To say it differently,
a quasi-static sequence is a series of equilibrium states. The entropy
for each equilibrium state is naturally a constant.
However, for each equilibrium state the
constant value of the entropy is different. This is what accounts
for an irreversible growth of entropy even in quasi-static processes.
Since in the analysis of Ref. \cite{DaFoPa} the entropy of the shell was
not considered, the growth of entropy in a quasi-static sequence of shell
positions could not be addressed.

The  explicit amount of  entropy (which remains constant for the
quasi-static reversible process mentioned above or increases for the
quasi-static irreversible  processes discussed later) depends on the
model description of the matter in the shell. This would determine the
function $b(r_+)$ in (\ref{egral}) and  the empirical coefficients $a$
and $\eta$ in (\ref{Sresult}).  In previous sections we saw how 
physical considerations of a general character severely restrict the 
range of the  coefficient $a$.
It is not so easy to constrain the range of $\eta$ without a 
precise description of the matter fields. In particular, it is not
difficult to see that one could account for most of the black hole
entropy if the value of $\eta$ is of order  $(r_{+}/l_p)^{(1-a)}$. If we
assume that $\eta$ is of order one, we obtain  the fundamental equation 
(\ref{Sfinal}) for a stable shell. For  fixed $r_+$, the entropy remains
constant as one lowers  the shell, and the maximum possible constant
value of the entropy  is attained for $a= 12/19$. 
Therefore for a stable shell at constant $r_+$ \begin{equation}
S \leq S_{max} = 
\pi \bigg( {{r_+}\over{{l_p}}} \bigg)^{(31/19)} = 
\bigg( {{l_p}\over{{r_+}}} \bigg)^{(7/19)} \, S_{BH} \ ,
\end{equation}
where $S_{BH}$ refers to the entropy of the black hole whose
ADM mass equals the ADM mass $m$ of the shell. 
In other words, the entropy of a black hole exceeds the
entropy of a stable shell of the same ADM mass by (at least) a factor of
order  $({r_+}/{{l_p}})^{(7/19)}$ whenever the thermal
equation of state for the latter takes the form (\ref{powerlaw}) and
$\eta$ is  of order one 
(namely $\eta = \pi (a +1)$).  
If the third law were to be enforced in the manner discussed in Section
III, the maximum entropy would occur for $a=0$. In this case, the entropy
of a black hole of radius $r_+$ would exceed the entropy of a stable
shell of the same ADM mass by a factor of order 
$({r_+}/{{l_p}})$.  In any case, in this simple example the
entropy of a shell of given ADM mass  would equal the entropy of
a black hole of the same mass only if the size of the horizon is of the
order of the Planck length. We stress that these estimates are based 
on the particular form (\ref{powerlaw}) of the thermal equation of state
whose overall multiplicative coefficient $\eta$ is assumed to be of 
order one.

Irreversible quasi-static processes characterized by different 
types of boundary data can be easily constructed. For given boundary 
conditions defining the process, the sequence of 
equilibrium states $(M, R)$ is dictated by Eq. (\ref{M}) and the
associated entropy at each state by (\ref{Sresult}).  
As an example, consider a quasi-static  process  obtained by 
constraining the position of the shell to be $R = R_0 = {\rm const.}$ 
In thermodynamical language, this is equivalent to assuming internal
constraints in the  system that induce the shell to be
restrictive with  respect to size.   The incoming energy
$dm \geq 0$ to the system  is fully spent in  increasing the 
energy $M$ (according to  Eq. (\ref{M}) with $R=R_0$) and none
transforms into mechanical work.  A possible  realization of this
kind of quasi-static process is the following: start with  an idealized
shell whose mass is small. As energy flows  into the system, the
local mass $M$ grows  quasi-statically until $r_+ \to R_0$, at which
limit we assume that  a black hole forms. Its entropy at each
equilibrium stage is given by (\ref{Sresult}). However, as we mentioned
before, we expect that this expression becomes  
inaccurate for values $r_+ \geq (24/25) R_0$. Therefore, for fixed values
of $\eta$ and $a$  (and $S_0 = 0$)  the  entropy attainable by
this  quasi-static process is  
\begin{equation} 
S \leq S_{max} =
{{\eta}\over{(a+1)}}  \bigg( {{24}\over{25}} {{R_0}\over{{l_p}}}
\bigg)^{(a+1)}  \ .
\end{equation}
If one assumes $\eta = \pi (a+1)$, the maximum entropy for a stable shell 
can be obtained from (\ref{Sfinal}) with $a=12/19$ and equals
\begin{equation}
S_{max} = \bigg( {{24}\over{25}} \bigg)^{(31/19)} \,
\bigg( {{l_p}\over{{R_0}}} \bigg)^{(7/19)} \, \, S_{BH} \,  
\approx \,  0.94  \bigg( {{l_p}\over{{R_0}}} \bigg)^{(7/19)} \, \, S_{BH}
\ , \end{equation}
where $S_{BH} = \pi (R_0/{l_p})^2$ is the entropy of the black hole
formed as $r_+ \to R_0$.

The previous analysis can be easily generalized to a wide variety of
processes involving quasi-static interchange of energy and work between 
the shell and a reservoir. 
We do not claim all of these processes to be physical: 
thermodynamics does not guarantee dynamics. 
They are simply not forbidden by thermodynamical arguments and 
provide interesting examples which illustrate the quasi-static behavior 
of entropy.

\section{Scaling and Gibbs-Duhem}

The integrated form of the first law for our system is
 \begin{equation}
	M = (a+1)\, T S - 2 p \, A - (a+1)\, T \, S_0 \ . \label{euler}
\end{equation}
This is easy to verify using  Eqs. (\ref{p}), (\ref{beta}) and
(\ref{powerlaw}) for the intensive parameters $(\beta, p)$ and Eq.
(\ref{Sresult}) for the entropy $S$. 
This ``Euler relation" implies that the entropy $S(M,A,N)$ is a
homogeneous  function of degree $(a+1)$ in $M$ and of degree $(a+1)/2$ in
$A$.  (Alternatively, the energy $M(S,A,N)$ is a homogeneous function  of
degree $1/(a +1)$ in $S$ and of degree one-half in $A$.)  
The scaling laws for the self-gravitating system are therefore: 
$M \to \lambda M$ ($r_+ \to \lambda r_+$), $A \to {\lambda}^2
A$  ($R \to \lambda R$), and $S \to {\lambda}^{(a+1)} S$. 
This behavior is due to the 
fact that the intensive variables have to be rescaled according to
$\beta \to \lambda^{a} \beta$, and 
$p \to {\lambda}^{-1} p$.

The Euler relation (\ref{euler}) illustrates that the scaling
laws characteristic of self-gravitating matter systems 
are different from the ones of ordinary
flat-spacetime thermodynamics.  We stress that this is so even if 
no black hole is present in the system.
Observe that in the limit $a = 0$ one does not recover the 
ordinary scaling:
the work term in (\ref{euler}) does not reduce to the familiar form  
$-p V$. This is partly a consequence of the role played by  area 
as the variable measuring the ``size" of a system. 
The difference between the scaling laws of standard thermodynamics  and
the ones characteristic of a black hole  has been recognized in
Ref.\cite{Yo}.  The expressions for a black hole can be recovered from
the above ones by taking the limit $a=1$ and $\eta = 2 \pi$.  As pointed
out in Ref. \cite{Yo} in the context  of  black holes, these scaling
properties  must play an important role in the description of phase
transitions involving self-gravitating systems. 

In standard thermodynamics there exists a relationship 
(the Gibbs-Duhem relation) among the various 
intensive parameters which is a consequence of the homogeneous 
first order degree of the fundamental relation \cite{Ca}.
It can be obtained  by combining the Euler relation with the 
first law of thermodynamics and states that the sum of products of 
the extensive parameters and the differentials of the corresponding 
intensive parameters vanishes, namely   $SdT - V dP + N d\mu =0$ in the
energy representation. The quantity  $V$ denotes the size of the
system and $\mu$ the chemical  potential conjugate to $N$. Because of the
homogeneous properties  of Eq. (\ref{euler}) discussed above,
this form of the Gibbs-Duhem  relation is {\it not} valid for a
self-gravitating system.  By differentiating Eq. (\ref{euler}) and
combining the result with the first law (\ref{TdS}) one obtains instead
\begin{equation} 
(a+1) \, S \, dT - 2 \, A \, dp + (a+1) \, N \, d\mu + a \, dM +
(a-1) \, p \, dA = 0  \ . \label{GD}
\end{equation}
In the limit $a=1$ this relationship reduces
to the corresponding one for a black hole, namely 
\begin{equation}
	2S\, dT - 2 A \, dp + 2N \, d\mu + dM = 0 \ .
\end{equation}

\section{Concluding remarks}

We have attempted to clarify in this paper the features of the 
fundamental thermodynamical equation of a matter
system which  arise {\it solely} from its self-gravitating 
character and are therefore independent of the  
microscopic structure of the matter fields. 
The fundamental equation  (\ref{egral}) is a consequence of the 
pressure equation of state (\ref{p}), the integrability of
the first law of thermodynamics, and the assumption of constant number
$N$. The pressure equation is in turn a direct consequence of the
gravitational junction conditions  (and
therefore of Einstein equations) at the position of a
two-dimensional surface  in  thermal and mechanical  equilibrium with
itself.  To  specify  the
fundamental equation completely one  needs to add  a phenomenological or
quantum description of the matter fields. Different models
 provide different functions $b(r_+)$.  Our purpose has been
not to classify the latter  (since this does not concern gravity)
but illustrate the method with a simple physical choice.

The methods adopted in this paper to explore the fundamental equation
of a self-gravitating system are closely related to methods of finding
fundamental equations in ordinary thermodynamics. Our phenomenological
description of the gravitational field through Einstein equations
provided us with an ``empirical" pressure equation of state, in much the
same spirit as one obtains, for example,  the van der Walls
pressure equation of state in fluid mechanics \cite{Ca}. 
A thermal equation of state has
then to be found by adopting the simplest expression that is  physically
reasonable and guarantees integrability for the entropy. 
In both phenomenological descriptions of simple fluids and
self-gravitating bodies, empirical coefficients have  to be assumed. 
The phenomenological approach can be
pushed forward by requiring the entropy to satisfy several physical
conditions which have thermodynamical origin.
 Ultimately, their precise values as well as
the range of applicability of the resulting fundamental equation  have to
be determined by experiment or by a fully statistical description of the
interactions. 
It would be very interesting to construct simple
 but realistic
models (involving, for example, scalar  fields) of the structure of
the shell. They may provide explicit values for the coefficients
as well as define the regions in which the quasi-static processes 
discussed in Section IV are a good approximation to the full
dynamical collapse. 
In any event, the fundamental thermodynamical equations arrived at are 
very rich and illustrate many of the results and strengths of a 
thermodynamical approach to quantum-statistical gravitational systems.

A black hole contained inside a spherical boundary in thermal 
equilibrium with a heat bath and obeying the semiclassical 
Hawking's  equation  (\ref{betabh})  
is not intrinsically stable.  
Therefore, it is of interest to investigate slightly more general 
thermal equations of state which would allow black hole 
 stability without altering 
the pressure equation of state (\ref{pbh}). 
This approach may in turn  provide some guidance into thermal
equations of state  which might be considered  as `corrections' to 
Hawking's equation and which might effectively incorporate  back-reaction 
and higher order effects. 
The results of this paper apply to a black hole located inside a spherical
cavity  whose thermal equation
of state is  generally given by Eqs.  (\ref{beta}) and (\ref{powerlaw}).
(Hawking's formula is recovered formally by taking the
limits $a=1$ and  $\eta = 2 \pi$.) 
The stability analysis of Section III can be
easily adapted to this type of black holes. 
It implies that values of $a$ smaller than unity in the thermal equation
of state
do guarantee stability for (finite) ranges of the boundary surface radii
larger than the  horizon radius.   For example, if $a=0.9$ the
black hole could be stable provided the  boundary radius satisfies $r_0
\leq 1.003\, r_+$. The range of values for $r_0$ that guarantees
stability increases  as $a$ decreases.  Of course, in a phenomenological
description like the one presented here there is no reason to select a
particular value of this coefficient.  In any case,  this would imply
a black hole entropy given by  Eq. (\ref{Sfinal}) where the exponent of
$r_+$ is  smaller than two.  It would be of interest to see whether
alternative expressions for the entropy of a black hole,  obtained  by 
including either higher-loop terms or different types of ``hair" in the
gravitational action  (see for example Ref. \cite{Vi} and 
references therein) could be expressed (at least partially) in the 
above mentioned form. We will return to this issue elsewhere.

We have focused our attention on the thermodynamics of equilibrium states
and therefore  only considered quasi-static 
(and not dynamical) processes.  
Whereas the local intrinsic stability conditions studied here are 
the  weakest stability restrictions one can impose to a system based
solely on thermodynamics, dynamical stability may
impose further restrictions. Mechanical stability under
dynamical perturbations of a shell surrounding a black hole has been 
studied recently in Ref. \cite{BrLoPo} by examining the 
equations of motion in the neighborhood of equilibrium 
configurations.  This analysis did not include 
a thermal behavior for a shell but found 
that, in the particular case  when no black hole is present, 
the largest region for  mechanical stability for a ``stiff" shell (for
which the speed of sound  equals the speed of light) occurs when $k \leq
0.395 $, or equivalently, for a radius $R$ larger than approximately
$1.185 \, r_+$. The critical value for stability might indeed be larger 
for smaller values of the speed of sound \cite{BrLoPo}. 
It is not difficult to see that this mechanical stability 
condition, if applied literally to our equilibrium 
states with a thermal equation of state given by (\ref{powerlaw})
would further restrict the value of the parameter $a$ to be smaller or 
equal than approximately  $0.37$. 
It would be of interest to generalize the dynamical analysis of Ref. 
\cite{BrLoPo} to incorporate the thermal behavior of the shell 
discussed in this paper.

\acknowledgments

It is a pleasure to thank Werner Israel for many stimulating 
discussions and critical remarks. 
The author is also indebted to Abhay Ashtekar and Lee Smolin  for
encouragement and useful conversations.  
Research support was received from the Natural Sciences and Engineering
Research Council of Canada, the National Science Foundation Grant 
PHY 93-96246,  and from the Eberly Research Funds of The Pennsylvania
State University.


\newpage


\begin{table}
\begin{center}
\caption{Intrinsic stability conditions for the fundamental equation 
(2.18). The table shows the range of $k$ in which the conditions are 
satisfied for different values of the parameter $a$. 
The symbol  ``$\surd $\," indicates that a criterion is satisfied in
the full physical range $0 \leq k \leq 1$, whereas the symbol 
``$\times $\," indicates that the criterion is {\it not} satisfied 
in this range.}
\vspace{4mm}
\begin{tabular}{|c||c|c|c|}
\hline
&&&\\
RANGE OF a&
$({{{\partial }^2 S}\over{\partial M^2}}) \leq 0$&
$({{{\partial }^2 S}\over{\partial A^2}}) \leq 0$&
$({{{\partial }^2 S} \over{\partial M^2}})({{{\partial }^2 S} \over
{\partial A^2}}) - 
({{{\partial}^2 S}\over{\partial A \partial M}})^2 \geq 0$\\
&&&\\
\hline\hline
&&&\\
$a \geq 3$&
$k \leq \sqrt{{1}\over{2a+1}}$&
$k \geq 1 - {{6}\over{a+3}}$&
$\times$ \\
&&&\\
\hline
&&&\\
$3 \geq a \geq 1$&
$k \leq \sqrt{{1}\over{2a+1}}$&
$\surd$&
$\times$, but (a) \\ 
&&&\\
\hline
&&&\\
$1 > a > {12 \over{19}} $&
$k \leq \sqrt{{1}\over{2a+1}}$&
$\surd$&
(b)\\
&&&\\
\hline
&&&\\
${12 \over{19}} \geq a > 0 $&
$k \leq \sqrt{{1}\over{2a+1}}$&
$\surd$&
(c)\\
&&&\\
\hline
&&&\\
$0 \geq a$&
$\surd$&
$\surd$&
$\surd$ \\
&&&\\
\hline
\end{tabular}
\end{center}
(a) For a=1, this condition is satisfied if $k=0$.\\
(b) $k$ is a solution of Eq. (3.7), but remains smaller than
$1/5$.\\
(c) $k$ is a solution of Eq. (3.7) always larger or
equal (for $a=12/19$) than $1/5$. See main text.   
\end{table}


\begin{references}

\bibitem{Ja} T. Jacobson, Phys. Rev. Lett. {\bf 75}, 1260 (1995).

\bibitem{GiHa} G. W. Gibbons and S. W. Hawking, Phys. Rev. {\bf D 15}, 
2738 (1977); {\bf D 15} 2752 (1977).

\bibitem{Pe} R. Penrose, in {\it General Relativity: an Einstein
Centenary Survey}, edited by S. W. Hawking and W. Israel (Cambridge
University Press, Cambridge, 1979).

\bibitem{DaFoPa} P. C. W. Davies, L. H. Ford, and D. N. Page, Phys. Rev. 
{\bf D 34}, 1700 (1986).

\bibitem{MaYo} E. A. Martinez and J. W. York, Jr., Phys. Rev. {\bf D 40},
2124 (1989).

\bibitem{Hi} W. A. Hiscock, Phys. Rev. {\bf D 40}, 1336 (1989).

\bibitem{Ca} H. B. Callen, {\it Thermodynamics and an Introduction to
Thermostatistics}, Wiley, New York, 1985.

\bibitem{BrMaYo} J. D. Brown, E. A. Martinez, and J. W. York, Jr., Phys.
Rev. Lett., {\bf 66}, 2281 (1991); in {\it Nonlinear Problems in
Relativity and Cosmology}, edited by J. R. Buchler, S. L. Detweiler, and
J.R. Ipser (New York Academy of Sciences, New York, 1991).

\bibitem{BrYoRev} J. D. Brown and J. W. York, Jr.,  ``The path integral
formulation of gravitational thermodynamics",  preprint IFP-UNC-491,
TAR-UNC-043, CTMP/007/NCSU, gr-qc/9405024.

\bibitem{Is} W. Israel, Nuovo Cimento {\bf 44B}, 1 (1966); 
{\bf 48B}, 463 (1967); V. de la Cruz and W. Israel, Nuovo
Cimento {\bf 51A}, 774 (1967).

\bibitem{La} C. Lanczos, Phys. Z. {\bf 23}, 539 (1922); Ann. Phys. 
(Germany) {\bf 74}, 518 (1924).

\bibitem{Yo} J. W. York, Jr., Phys. Rev. {\bf D 33}, 2092 (1986).

\bibitem{HaEl} S. W. Hawking and G. F. R. Ellis, {\it The Large Scale
Structure of Space-time}, Cambridge University Press, Cambridge, 1973.

\bibitem{FrHoKo} J. Frauendiener, C. Hoenselaers, and W. Konrad, Class.
Quantum Grav. {bf 7}, 585 (1990). 

\bibitem{To} R. C. Tolman,{\it Relativity, Thermodynamics, and 
Cosmology}, Oxford University Press, Oxford, 1935.

\bibitem{BrYo1} J. D. Brown and J. W. York, Jr., Phys. Rev. {\bf D47},
1407 (1993).

\bibitem{Ha} S. W. Hawking, Comm. Math. Phys. {\bf 43}, 199 (1975).

\bibitem{ensembles} J. D. Brown, G. L. Comer, E. A. Martinez, J.
Melmed, B. F. Whiting, and J. W. York, Jr., Class. Quantum Grav. {\bf
7}, 1433 (1990); H. Braden, J. D. Brown, B. F. Whiting, and J. W. York,
Jr., Phys. Rev. {\bf D 42}, 3376 (1990).

\bibitem{UnWa} W. G. Unruh and R. M. Wald, Phys. Rev. {\bf D 25}, 942 
(1982); {\bf D 27}, 2271 (1983).

\bibitem{Be81} J. D. Bekenstein, Phys. Rev. {\bf D 23}, 287 (1981). 

\bibitem{Be94} J. D. Bekenstein, Phys. Rev. {\bf D 49}, 1912 (1994), and
references cited therein. 

\bibitem{Be73} J. D. Bekenstein, Phys. Rev. {\bf D 7}, 2333 (1973).

\bibitem{Wa} R. M. Wald, {\it General Relativity}, The University of
Chicago Press,  Chicago, 1984.

\bibitem{Vi} M. Visser, Phys. Rev. {\bf D 48}, 583 (1993).

\bibitem{BrLoPo} P. R. Brady, J. Louko, E. Poisson, Phys. Rev. 
{\bf D 44}, 1891 (1991).

\end{references}
\end{document}